\documentstyle[prl,aps]{revtex}
\begin{document}

\tolerance=10000

% New control sequences
\newcommand{\be}{\begin{equation}}
\newcommand{\ee}{\end{equation}}
\newcommand{\bea}{\begin{eqnarray}}
\newcommand{\eea}{\end{eqnarray}}

\twocolumn[\hsize\textwidth\columnwidth\hsize
     \csname @twocolumnfalse\endcsname

\title{Interaction Effect in the Kondo Energy of the Periodic 
Anderson--Hubbard Model}
\author{K. Itai and P. Fazekas\\
Research Institute for Solid State Physics,\\
 P.O. Box 49, Budapest 114,  H--1525 Hungary}

\maketitle

\begin{abstract}
We extend the periodic Anderson model by switching on a Hubbard $U$ for the  
conduction band. The nearly integral valent limit of the 
Anderson--Hubbard model is studied with the Gutzwiller variational method. 
The lattice Kondo energy shows $U$-dependence both in the prefactor and the 
exponent. Switching on $U$ reduces the Kondo scale, which can be understood 
to result from the blocking of hybridization. At half-filling, we find a 
Brinkman--Rice-type transition from a Kondo insulator to a Mott insulator. 
Our findings should be relevant for a number of correlated two-band models 
of recent interest.
\end{abstract}

\vskip2pc]
\newpage

The usual scenario of the Kondo effect (either for a single impurity, or 
for the Kondo lattice) envisages that the localized moments of strongly 
correlated electrons are spin-compensated by a non-interacting electron gas. 
Recently, there has been a great deal of interest in the more difficult 
problem when the spin-compensating medium itself is a more-or-less strongly 
correlated electron system characterized by its own Hubbard $U$. Attention 
has been concentrated on the impurity case \cite{SF}, but there are 
also preliminary results on periodic models such as the Kondo--Hubbard lattice 
model \cite{shi}. In all studies, one finds a strong $U$-dependence of the 
Kondo energy. The subsystem described by the Hubbard model can be replaced 
by other correlated models such as the $t$--$J$ model \cite{MCC}, the 
quantum Heisenberg model \cite{ITKF}, or a Luttinger liquid \cite{Li}.

The standard Kondo problem can be treated at two levels: one can start either  
from the purely fermionic Anderson model, or from the Kondo Hamiltonian which 
is the large-$U$ effective Hamiltonian of the Anderson model. In the relevant 
parameter range the two approaches lead to equivalent results. Remarkably, 
once the Hubbard $U$ of the conduction electron sea is switched on, it 
matters very much whether the impurity is described by the Anderson or the 
Kondo model: one finds a reduced Kondo coupling for the Anderson impurity 
\cite{SF}, while the Kondo energy is enhanced if the Kondo Hamiltonian is 
used \cite{HF}. For an Anderson impurity in a Luttinger liquid, one even finds 
a non-monotonic $U$-dependence, with the Kondo temperature increasing 
at weak $U$, and decreasing at strong $U$ \cite{Li}. These findings are not 
necessarily mysterious: the handwaving argument might be that 
in the Anderson case, switching on $U$ blocks hybridization and therefore 
decreases the effective Kondo coupling. In contrast, in the Kondo Hamiltonian 
the coupling is already given, and then switching on $U$ helps the singlet 
formation because a more correlated electron gas offers more uncompensated 
spins. It appears that the Kondo impurity model with a Hubbard band 
is not the effective Hamiltonian for an Anderson impurity in the same 
Hubbard band, and it is worth to explore the two problems independently. 
The same holds for periodic models where it depends on the physical nature of 
the system whether the Anderson--Hubbard lattice model, or a Kondo--Hubbard 
lattice model is a better starting point. We note 
the suggestion \cite{FD} that the curious correlated semiconductor FeSi can 
be described by an Anderson--Hubbard-type model in which the spin-compensating 
and spin-compensated electrons play symmetrical roles. Perhaps a Kondo-type 
Hamiltonian gives a better starting point for the Nd$_{2-x}$Ce$_x$CuO$_4$ 
system \cite{FZZ,IMF}.    
  
We consider the periodic Anderson--Hubbard model which describes an array of 
strongly correlated $f$-sites hybridized with a moderately strongly 
interacting $d$-band:
\bea
{\cal H} & = &  \sum_{{\bf k},\sigma}\epsilon_d({\bf k})
d_{{\bf k}\sigma}^{\dagger}d_{{\bf k}\sigma}  
+ \epsilon_f\sum_{{\bf j},\sigma}{\hat n}_{{\bf j}\sigma}^f
+ U_f\sum_{\bf j}{\hat n}_{{\bf j}\uparrow}^f
{\hat n}_{{\bf j}\downarrow}^f 
\nonumber \\[2mm]
& & +U_d\sum_{\bf j}{\hat n}_{{\bf j}\uparrow}^d{\hat n}_{{\bf j}\downarrow}^d
-v\sum_{{\bf j},\sigma}(f_{{\bf j}\sigma}^{\dagger}
d_{{\bf j}\sigma} + d_{{\bf j}\sigma}^{\dagger}f_{{\bf j}\sigma})
\label{eq:FD}
\eea
where ${\hat n}_{{\bf j}\sigma}^f=f_{{\bf j}\sigma}^{\dagger}
f_{{\bf j}\sigma}$, etc., the {\bf k} are wave vectors, and the {\bf j} are 
site indices. The $d$-bandwidth is $W$. In what follows, we take the 
strongly asymmetric Anderson model with $U_f\to\infty$ and the $f$-level 
$\epsilon_f<0$ sufficiently deep-lying so that we are in the Kondo limit: 
$1-n_f\ll 1$ where the $f$-valence is defined as 
$n_f=\langle\sum_{\sigma}{\hat n}_{{\bf j}\sigma}^f\rangle$.  The total 
electron density (per site, for one spin) is 
$n=\langle\sum_{{\bf j},\sigma}{\hat n}_{{\bf j}\sigma}^f+
\sum_{{\bf j},\sigma}{\hat n}_{{\bf j}\sigma}^d\rangle/2L$, where $L$ is the 
number of lattice sites. We will assume $1/2\le n\le 1$, so that there are 
enough electrons to fill at least the $f$-levels, and the $d$-band filling is 
variable up to half-filling. 

We use the Gutzwiller variational method, generalizing a previous treatment 
of the periodic Anderson model \cite{FB}. Limitation of space preventing us 
from going into details, we just mention a few salient points. The trial 
state is obtained by Gutzwiller-projecting the optimized hybridized-band state 
\be
|\Psi\rangle  =  {\hat P}_{\rm G}^d\cdot{\hat P}_{\rm G}^f\cdot
\prod_{\bf k} \prod_{\sigma} [u_{\bf k}
f_{{\bf k}\sigma}^{\dagger} + v_{\bf k}d_{{\bf k}\sigma}^{\dagger}]|0\rangle
\label{eq:psi1}
\ee
where the mixing amplitudes $u_{\bf k}/v_{\bf k}$ are treated as independent 
variational parameters. The Gutzwiller projector for the 
$d$-electrons is
\be
{\hat P}_{\rm G}^d = \prod_{\bf g}[1-(1-\eta){\hat n}_{{\bf g}\uparrow}^d
{\hat n}_{{\bf g}\downarrow}^d]
\ee
where the variational parameter $\eta$ is controlled by $U_d$. For the 
$f$-electrons the full Gutzwiller projection is taken
\be
{\hat P}_{\rm G}^f = 
\prod_{\bf g}[1-{\hat n}_{{\bf g}\uparrow}^f{\hat n}_{{\bf g}\downarrow}^f].
\ee
Here we consider only non-magnetic solutions corresponding to a mass-enhanced 
metal at $n<1$, or a renormalized-hybridization-gap insulator at $n=1$.
The study of magnetic ordering, especially in the less-than-half-filled case, 
should prove interesting: metallic ferromagnetism has been found in the 
Kondo--$t$--$J$ model \cite{MCC}, as well as in our preliminary investigation 
\cite{IF2} of the Fu--Doniach model \cite{FD}. We note that Tasaki's 
two-band model for which (for a certain choice of the parameters) the 
existence of ferromagnetism can be shown exactly \cite{Tas}, is also of the 
form (\ref{eq:FD}) but with a dispersive $f$-band and a {\bf k}-dependent 
hybridization.  
 
The lengthy details of the optimization procedure, along with results for the 
behaviour away from the nearly integral valent (Kondo) limit, will be 
published elsewhere \cite{IF2}. Using the Gutzwiller approximation, and 
minimizing with respect to the mixing amplitudes, we find an expression for 
the ground state energy density ${\cal E}$, which describes a renormalized 
$d$-band hybridized with effective $f$-levels:
\bea
{\cal E} & =  & \frac{1}{L}\sum_{{\bf k}\in {\rm FS}} \left[ 
q_d\epsilon_d({\bf k}) +{\tilde\epsilon}_f - \sqrt{(q_d\epsilon_d({\bf k}) 
-{\tilde\epsilon}_f)^2+4{\tilde v}^2}\right]
\nonumber \\[2mm]
& & +(\epsilon_f-{\tilde\epsilon}_f)n_f + U_d\nu_d
\label{eq:E}
\eea
where the {\bf k}-sum extends over the $U_f=U_d=0$ Fermi sea (in a manner 
familiar from the Anderson lattice case \cite{FB}, the Gutzwiller method 
respects Luttinger's theorem and leaves the Fermi volume unchanged). The 
$d$-band kinetic energy renormalization factor $q_d$ is the same as in the 
Gutzwiller treatment \cite{G} of a Hubbard model with filling $n_d=n-n_f/2$. 
Since we assumed a dispersionless $f$-band, the $q$-factor for the 
$f$-electrons $q_f=(1-n_f)/(1-n_f/2)$ 
appears only in the renormalized hybridization amplitude 
${\tilde v}=\sqrt{q_dq_f}v$. The renormalized band structure contains the 
effective $f$-level ${\tilde\epsilon}_f$; this shift is compensated by the 
last-but-one term of ${\cal E}$. The last term gives the $d$--$d$ interaction 
energy: it contains $\nu_d$, the density of doubly occupied $d$-orbitals. 
${\cal E}$ still has to be optimized with respect to $\nu_d$. 

Henceforth we assume a band with a constant density of states 
$\rho(\epsilon)=1/W$, lying in the interval $\epsilon\in [-W/2,W/2]$. 
Furthermore, we restrict ourselves to the weak hybridization regime $v\ll W$, 
where a number of physical quantities shows an exponential Kondo-like 
behaviour. One of them is the small deviation from integral $f$-valence. 
Minimizing with respect to $\nu_d$ requires
\be
\frac{U_d}{W}+\left[-\frac{1}{4}+\left(\frac{n_f}{2}-n_d^0\right)^2
-4\left(\frac{v}{W}\right)^2\frac{1-n_f}{q_d}\right]
\frac{\partial q_d}{\partial \nu_d}=0
\ee
which couples the non-integral valence to the Hubbard $U_d$. The solution is
\be
1-n_f = \frac{n_d^0\; q_d^0}{4(v/W)^2}\cdot \exp{\left\{-\frac{\mu_0(U_d)-
\epsilon_f}{4(v^2/W)}\right\} }.
\label{eq:1mnf}
\ee
In the expression for the optimized total energy density
\bea
{\cal E} & = & \epsilon_f -n_d^0(1-n_d^0)q_d^0W + U_d\nu_d^0
\nonumber \\[2mm]
& & -Wn_d^0\; q_d^0\cdot \exp{\left\{-\frac{\mu_0(U_d)-
\epsilon_f}{4(v^2/W)}\right\} }
\label{eq:ekon}
\eea
the first line gives the energies of decoupled $f$-, and $d$-electrons. The 
coupling between the two subsystems is described by the last term which we 
identify as the Kondo energy of the Anderson--Hubbard model. It turns out 
to be proportional to $1-n_f$. An analogous relationship between the valency 
and the Kondo temperature holds in the single impurity case \cite{Ram}.

In the above equations, $n_d^0=n-1/2$ is the $v=0$ value of the conduction 
electron density (per spin). $q_d^0$ is the $q$-factor taken with 
$n_d^0$, and $\mu_0(U_d)$ 
is the chemical potential of the Hubbard subsystem calculated (in the 
Gutzwiller approximation) for band filling $n_d^0$.  

(\ref{eq:1mnf}) and (\ref{eq:ekon}) are formal relationships in the sense that 
they express the solution in terms of the optimized parameters of the Hubbard 
subsystem, which have no simple closed form for arbitrary $U_d$ away from
half-filling. We will give the results of small-$U_d$, and large-$U_d$, 
expansions. First, however, let us discuss the formal solution.

(\ref{eq:1mnf}) and (\ref{eq:ekon}) are the main results of the present paper. 
They say that 
switching on $U_d$ influences the characteristic Kondo scale in two ways: 
through the prefactor and the exponent. The prefactor describes the 
correlation-induced narrowing of the $d$-band. The exponential factor can 
be written as $\exp{(-W/J_{\rm eff})}$ where we introduced the effective Kondo 
coupling
\be
J_{\rm eff} = \frac{4v^2}{\mu_0(U_d)-\epsilon_f}
\label{eq:effj}
\ee
It differs from the Anderson lattice result \cite{RU,FB} inasmuch as $\mu_0(0)=
(n_d^0-1/2)W$ is replaced by the $U_d$-dependent chemical potential 
$\mu_0(U_d)$. 

Our $J_{\rm eff}$ differs from that of the corresponding impurity problem 
\cite{SF} in two respects. First, it has $4v^2$ in the numerator instead of 
$2v^2$: the ``lattice enhancement of the Kondo effect'' \cite{RU} works also 
in the present model. Second, the denominator is also different, as we will 
discuss shortly.

In the weak coupling limit ($U_d\ll W$), the standard Hubbard model 
results 
\be
\mu_0(U_d) \approx \mu_0(0)+n_d^0 U_d-n_d^0(1-n_d^0)(1-2n_d^0)(U_d^2/W)
\label{eq:wce1}
\ee
and 
\be
q_d \approx 1-[1-4(1-n)^2]\left(\frac{U_d}{2W}\right)^2.
\label{eq:wce2}
\ee
have to be replaced into (\ref{eq:1mnf}), or (\ref{eq:ekon}). We point it out 
that both in the 
prefactor and in the exponent, the appearance of $U_d$ acts to diminish the 
Kondo scale. 

It is worth to discuss the half-filled case $n=1$  which by 
Luttinger's theorem arguments belongs to a correlated non-magnetic insulator. 
The second-order term in (\ref{eq:wce1}) vanishes at $n_d=1/2$. 
In fact, at half-filling $\mu_0(U_d)=U_d/2$ holds for arbitrary $U_d$. (Note 
that in the single-impurity result \cite{SF}, $U_d$ appears instead of our 
$U_d/2$). (\ref{eq:wce2}) reduces to the wellknown \cite{BR} 
$q_d=1-(U_d/2W)^2$ which holds right up to $U_d=2W$. As $U_d\to 2W$ from 
below, the $d$-band undergoes a Brinkman--Rice transition, the system becomes 
integral valent ($n_f=1$) and the Kondo 
effect is completely quenched. Within the Gutzwiller-type treatment of 
correlations, it is not unexpected that this should happen: the Kondo effect 
is due to the $f$--$d$ hybridization which causes charge fluctuations in the 
$d$-shells. In the Brinkman--Rice scenario of the Mott transition, these 
fluctuations get completely suppressed, and then so is the Kondo effect.
--- We should emphasize, though, that while the usual one-band 
Brinkman--Rice transition is a metal--insulator transition, here we are 
speaking about an insulator--insulator transition. For $U_d<2W$, the 
effective $d$-band has a finite width and the renormalization of the 
hybridization matrix element gives rise to a modified Kondo gap. Calculating 
it requires a straightforward extension of the variational method to the 
$n>1$ case. Having done this, the gap can be calculated as the discontinuity 
of the chemical potential:
\bea
\Delta & = & \left(\frac{ d{\cal E}}{2dn}\right)_{n=1+0}-
\left(\frac{ d{\cal E}}{2dn}\right)_{n=1-0}
\nonumber \\[2mm]
& = & W\left(1-\frac{U_d^2}{4W^2}\right)\cdot\exp{\left\{- 
\frac{\frac{U_d}{2}-\epsilon_f}{4(v^2/W)}\right\} }
\eea
The result can be rewritten as $\Delta=2q_fv^2/W$, i.e., $\Delta$ can be 
interpreted as the renormalized hybridization gap. --- For $U_d>2W$ all 
polarity fluctuations cease and the system becomes a Mott insulator.  

Returning to the case of a general $n_d^0\le 1/2$, in the strong coupling 
limit ($U_d\gg W$) we replace 
\be
\mu_0(U_d)\approx \mu_0(0)+Wn_d^0 -n_d^0(1-3n_d^0)\frac{W^2}{U_d}
\label{eq:sc1}
\ee
and
\be
q_d^0 \approx \frac{1-2n_d^0}{1-n_d^0}\left(1+2n_d^0\frac{W}{U_d}\right)
\label{eq:sc2}
\ee
into (\ref{eq:1mnf}) and (\ref{eq:ekon}). The Kondo energy vanishes both at 
$n_d^0\to 0$ (we run out of conduction electrons) and at $n_d^0\to 1/2$ 
(because we are above the Brinkman--Rice transition).

Comparing the $U_d\to\infty$ expression  
\be
1-n_f = \frac{n_d^0(1-2n_d^0)}{1-n_d^0}\cdot\frac{W^2}{4v^2}
\cdot\exp{\left\{ -\frac{\mu_0(0)+Wn_d^0-\epsilon_f}{4(v^2/W)}
\right\}}  
\label{eq:str}
\ee
to the $U_d=0$ (Anderson lattice) case, we see that again, both the prefactor 
and the new term in the exponent tend to reduce the Kondo scale. Taken 
together with the previous weak coupling result, we conclude that the 
effective Kondo coupling is less than its $U_d=0$ value for all $U_d>0$.
For $n$ slightly less than 1, the Anderson--Hubbard model predicts an 
even greater mass emhancement than the Anderson lattice, possibly accounting 
for findings like the heavy-fermion behaviour of Nd$_{2-x}$Ce$_x$CuO$_4$.

It is curious that the small $W/U_d$-correction seen in (\ref{eq:sc1}) changes 
sign as a function of the band filling $n_d^0$ at $n_d^0=1/3$. If it were 
for the exponent alone, it might have been expected that at $1/3<n_d^0<1/2$, 
at sufficiently large $U_d$, the Kondo scale increases towards its 
$U_d\to\infty$ value, and thus on the whole, its $U_d$-dependence in 
non-monotonic. However, our numerical results indicate that this tendency is 
counteracted by the stronger $U_d$-dependence of the prefactor and for 
reasonable parameters, the Kondo scale is a monotonically decreasing function 
of $U_d$. We should mention, though, that this aspect of the behaviour is 
strongly influenced by the assumption about the form of the density of 
states. --- All in all, our lattice result is qualitatively similar to the 
Anderson impurity results in the strong-coupling limit \cite{SF,Li}, while a 
discrepancy in the weak-coupling case \cite{Li} may be ascribable to the 
peculiarities of the Luttinger liquid. 

We should mention the limitations of our variational method. The use of 
the Gutzwiller approximation is justifiable only in the limit of infinite 
dimensionality but for many purposes, it should be a good approximation in 
three dimensions. In contrast, a number of previous studies dealt with a 
magnetic impurity imbedded in a one-dimensional system \cite{shi,ITKF,Li}. 
An essential change of physics in going from one to two dimensions has been 
pointed out in \cite{IMF}; three-dimensions should differ from both. --- 
As for the limitations of the Ansatz 
(\ref{eq:psi1}), it shares with similar treatments of the Anderson 
\cite{RU,FB} and Kondo \cite{ShF} lattices the shortcoming of neglecting 
intersite correlations other than those following from the Fermi statistics 
of the projected Fermi sea.   
 
To summarize, we found that switching on the Hubbard $U_d$ of the conduction 
electron sea tends to diminish the Kondo energy scale of the periodic 
Anderson model. This behaviour is just the opposite of that found for the 
Kondo--Hubbard lattice \cite{shi}. We believe that this discrepancy is 
understandable, 
and that the suppression of the Kondo scale in the Anderson--Hubbard lattice 
model arises from the fact that increasing $U_d$ suppresses charge 
fluctuations, and therefore blocks the hybridization processes which would 
lead to the Kondo effect. $U_d$ appears both in the exponent and in the 
prefactor. In the exponent, we find the $U_d$-dependent chemical potential of 
the $d$-band, while the prefactor reflects the correlation-induced narrowing 
of the conduction band, and it is sensitive to the nearness of the 
Brinkman--Rice transition. Including the effects of a finite $f$-bandwidth, 
similar 
conclusions should apply to the non-magnetic phases of such models as 
Tasaki's two-band model \cite{Tas}, or the Fu--Doniach model \cite{FD}. It 
appears to us that the common difficulty in many models of current interest 
is the description of Kondo compensation by a set of interacting electrons, 
and hence we hope that the relevance of the issues discussed here transcends 
the usual domain of $f$-electron physics.

P. F. gratefully acknowledges financial support by 
the Hungarian National Science Research Foundation grant OTKA T-014201. Part 
of this work was done during the authors' stay at the 
Tokyo Institute of Technology where they enjoyed the generous hospitality of 
H. Shiba. P.F. wishes to thank H. Shiba, N. Shibata, and C. Ishii for 
enlightening discussions.


\begin{thebibliography}{99}

\bibitem{SF} T. Schork and P. Fulde: Phys. Rev. B {\bf 50}, 1345 (1994).
\bibitem{shi} N. Shibata, K. Ueda, and C. Ishii: unpublished.
\bibitem{MCC} S. Moukouri, Liang Chen, and L.G. Caron: preprint.
\bibitem{ITKF} J. Igarashi, T. Tonegawa, M. Kaburagi, and P. Fulde: Phys. 
Rev. B {\bf 51}, 5814 (1995).
\bibitem{Li} Y.M. Li: Phys. Rev. B {\bf 52}, 6979 (1995).
\bibitem{HF} G. Khaliullin and P. Fulde: Phys. Rev. B {\bf 52}, 9514 (1995).
\bibitem{FD} C. Fu and S. Doniach: Phys. Rev. B {\bf 51}, 17439 (1995).
\bibitem{FZZ} P. Fulde, Z. Zevin, and G. Zwicknagl: Z. Phys. B {\bf 92}, 133 
(1993).
\bibitem{IMF} J. Igarashi, K. Murayama, and P. Fulde: Phys. Rev. B {\bf 52}, 
15966 (1995).
\bibitem{FB} P. Fazekas and B.H. Brandow: Phys. Scr. {\bf 36}, 809 (1987).
\bibitem{IF2} K. Itai and P. Fazekas: in preparation.
\bibitem{Tas} Hal Tasaki: Phys. Rev. Lett. {\bf 75}, 4678 (1995).
\bibitem{G} M.C. Gutzwiller: Phys. Rev. {\bf 137}, A1726 (1965). See also the 
review by D. Vollhardt: Rev. Mod. Phys. {\bf 56}, 99 (1984).
\bibitem{Ram} T.V. Ramakrishnan: J. Magn. Magn. Mater. {\bf 63}{\&}{\bf 64}, 
529 (1987).
\bibitem{RU} T.M. Rice and K. Ueda: Phys. Rev. B {\bf 36}, 6420 (1986).
\bibitem {BR} W.F. Brinkman and T.M. Rice: Phys. Rev. B {\bf 2}, 4302 (1970). 
\bibitem{ShF} H. Shiba and P. Fazekas: Progr. Theor. Phys. Suppl. {\bf 101}, 
403 (1990).
\end{thebibliography}
\end{document}